\documentclass{optica-article}
\journal{oe}
\articletype{Research Article}
\usepackage{upgreek}
\usepackage{lineno}
\usepackage{svg}
\usepackage{braket}
\usepackage{booktabs}
\newif\ifproofread
\newcommand{\changemarker}[1]{%
\ifproofread
\textcolor{red}{#1}%
\else
#1%
\fi
}

\begin{document}
\proofreadfalse
\title{Heterogeneous sapphire-supported low-loss photonics platform}

\author{Yubo Wang\authormark{1}, Yu Guo\authormark{1}, Yiyu Zhou\authormark{1}, Hao Xie\authormark{1}, and Hong X. Tang\authormark{1,*}}

\address{
\authormark{1}{Department of Electrical Engineering, Yale University, New Haven, CT 06520, USA}
}

\email{\authormark{*}hong.tang@yale.edu}


\begin{abstract}
Sapphire is a promising wideband substrate material for visible photonics. It is a common growth substrate for III-nitride light-emitting diodes and laser structures. Doped sapphires are important gain media foundational to the development of titanium-sapphire and ruby lasers. For lasers operating at visible and near-infrared wavelengths, a photonic platform that minimizes loss while maximizing gain material overlap is crucial. Here, we introduce a novel low-loss waveguiding strategy that establishes high-performance integrated photonics on sapphire substrates. This platform achieves a high intrinsic quality factor of 5.6 million near 780 nm and features direct compatibility with a range of solid-state laser gain media. 
\end{abstract}


\section{Introduction} 
Silicon photonics has gained significant success across various areas due to its ability to support high-density integration and its compatibility with silicon microelectronics fabrication processes, with its application in the telecom sector at the 1.5-$\mu$m wavelength serving as a prime example. Sapphire is considered an important wide bandgap material in photonics and is acclaimed for its use as important gain media such as titanium-doped sapphire and ruby, in addition to being a preferred substrate for blue semiconductor devices like gallium nitride emitters. The advent of visible photonic integrated circuits (PICs) is set to transform fields like quantum metrology\cite{niffenegger2020integrated}, \changemarker{optical} sensing\changemarker{\cite{degen2017quantum, mohanty2020reconfigurable, singh2023alanine}}, and consumer display optics, driven by progress in integrating visible to near-infrared (near-IR) lasers through both heterogeneous\cite{tran2022extending} and hybrid integration techniques\cite{siddharth2022near, franken2021hybrid}. The landscape of platforms for visible wavelengths encompasses silicon-supported silicon nitride\cite{sanna2024sin, poon2024silicon, corato2024absorption, ye2023foundry, xiang2022silicon, sinclair20201, pfeiffer2018ultra, buzaverov2023low}, alumina\cite{west2019low,he2023ultra}, and lithium niobate\cite{mckenna2020cryogenic, desiatov2019ultra}. Notably, on sapphire substrates, III-Nitride materials such as aluminum nitride and gallium nitride have exhibited remarkably low losses, reaching down to 5.3 dB/cm in the visible\cite{lu2018aluminum} and 8 dB/cm in the UV bands\cite{liu2018ultra}.

While a wide range of photonics platforms has been developed, there remains a notable gap in the capabilities of existing platforms for certain applications, such as heterogeneous laser integration and on-chip sensing\cite{zektzer2021nanoscale}, where a significant overlap with cladding is essential. High-confinement platforms, which contain most of the optical field within the waveguide, face challenges in achieving sufficient modal overlap with the gain material when integrated heterogeneously. On the other hand, low-confinement platforms like thin core silicon nitride\cite{puckett2021422, morin2021cmos, jin2021hertz, bauters2011planar} and tantala\cite{zhao2020low}, offer ultralow optical losses and notable cladding overlap, however require large bending radii, which can range from millimeters to centimeters. This affects radiation loss and device compactness, often necessitating reliance on foundry-based manufacturing processes. Additionally, the capabilities and potential applications of sapphire-based platforms have yet to be fully explored.

In this study, we demonstrate low-loss waveguides on sapphire substrates that effectively mitigate the trade-off between minimizing radiation loss and maximizing modal overlap, thus achieving a notably compact design. 
We achieved a quality factor of 5.6 million at near-visible wavelengths and substantial modal overlap with the top cladding of up to 35\%. This integrated circuit platform is scalable to shorter wavelengths\changemarker{\cite{wang2023photonic}}, capable of integrating additional functionalities such as thermal tuning elements, thus particularly well-suited for integrating sapphire-based solid-state lasers and the heterogeneous integration of III-N emitters\changemarker{\cite{holguin2022narrow, arefin2020iii}}.

\begin{figure}[h!]
\centering\includegraphics[width=\textwidth]{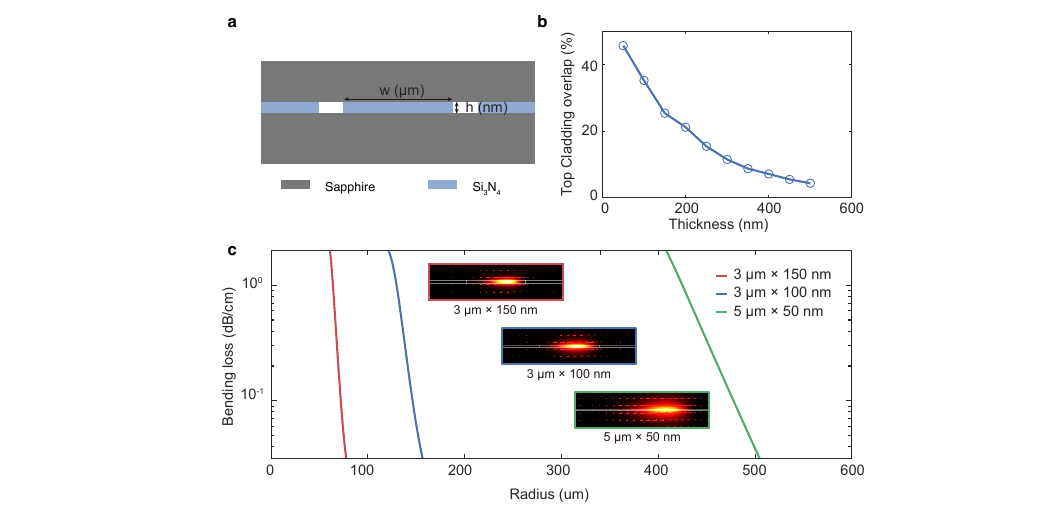}
\caption{(a) Schematic illustration of the proposed sapphire sandwich waveguide geometry. (b) Simulated top cladding overlap factor for the sapphire-based waveguide, where the waveguide width is 3 µm. (c) The waveguide bending loss as a function of the bending radius in three geometries: 3\,$\upmu$m$\,\times\,$150\,nm,  3\,$\upmu$m$\,\times\,$100\,nm, and  5\,$\upmu$m$\,\times\,$50\,nm. The insets show typical TE mode profiles corresponding to the three geometries at bending radii of 200 $\upmu$m, 400 $\upmu$m, and 800 $\upmu$m, from top to bottom, respectively. Arrow direction and length represent the local electric field's direction and strength, respectively. 
}
\label{fig1}
\end{figure}

\section{Design and Fabrication}

The waveguide discussed in this manuscript is composed of three layers: a bottom sapphire substrate, a silicon nitride (SiN) waveguide core, and a top sapphire cladding, as depicted in Figure \ref{fig1}(a). Among various design considerations for creating low-loss waveguides, selecting the appropriate thickness for the waveguide core is paramount. A thinner core can reduce scattering losses due to decreased modal confinement, albeit at the cost of a larger bending radius, which in turn, reduces the potential for high-density integration. Figure \ref{fig1}(b) illustrates the variation of the overlap factor with the top cladding across different waveguide thicknesses. The overlap factor is defined as $\Gamma_{\text{clad}} = \frac{1}{2} \sqrt{\frac{\epsilon_0}{\mu_0}}\frac{\int_{A_{\text{clad}}} n_{\text{clad}} |E|^2 \, dA}{\int_A \langle S_z \rangle \, dA}$, where $\epsilon_0$,$\mu_0$ are the vacuum permittivity and vacuum permeability, $n_{\text{clad}}$ is the refractive index of the cladding, $\langle S_z \rangle$ is the average Poynting vector, $A_{\text{clad}}$ and $A$ are the cladding integration and overall integration area. A thickness of 50 nm achieves an overlap factor of up to 46\%, fostering significant optical-ion interactions when the top cladding is doped. However, this benefit comes at a cost: thinner SiN waveguides incur higher bending losses. Figure \ref{fig1}(c) presents the simulated bending losses for waveguides of three thicknesses—50 nm, 100 nm, and 150 nm—with widths of 5 $\upmu$m for the 50 nm thickness and 3 $\upmu$m for the 100 nm and 150 nm thicknesses, specifically for the fundamental TE mode. Considering the limitations of the typical electron beam (ebeam) writing field of 1mm by 1mm for resonator fabrication, we selected a waveguide thickness of 100 nm for subsequent device fabrication.
\begin{figure*}[h!]
\centering
\includegraphics[width=1\textwidth]{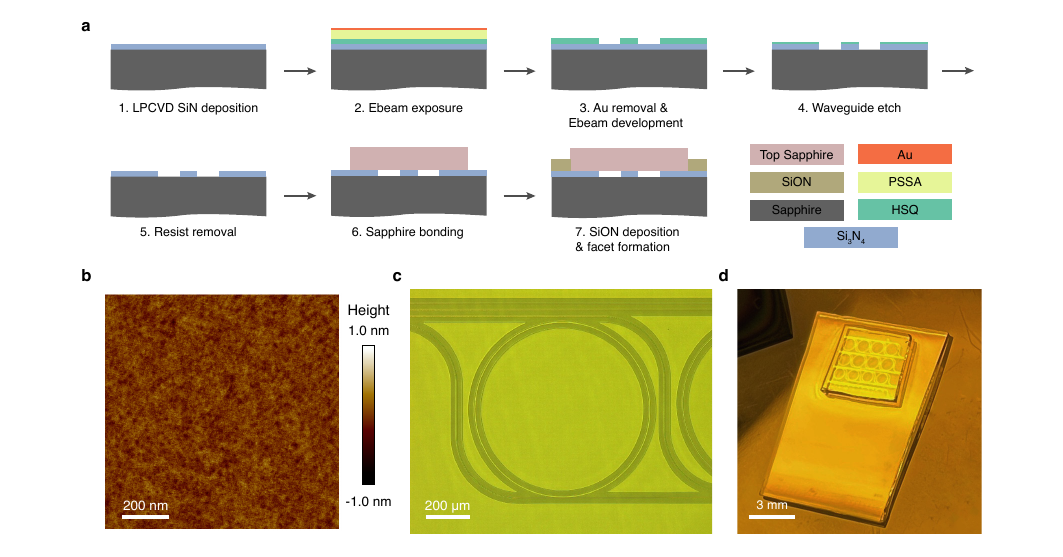}
\caption{\changemarker{(a)} Process flow for sapphire-based waveguide device fabrication, which includes steps from low-pressure chemical vapor deposition (LPCVD) of silicon nitride (SiN), etching of waveguides, bonding of the top sapphire layer, and the deposition of silicon oxynitride (SiON). (b) Atomic force microscopy image illustrating the sapphire substrate utilized for SiN growth, exhibiting a roughness average (Ra) of \changemarker{0.10} nm. (c) Microscopic image of a microring resonator. (d) Optical image of the final device, scale bar 3 mm.}
\label{fig2}
\end{figure*}

The fabrication sequence is outlined in Figure \ref{fig2}(a). The process begins with depositing a 100 nm thick layer of stoichiometric silicon nitride via low-pressure chemical vapor deposition (LPCVD) onto a sapphire substrate \changemarker{by Rogue Valley Microdevices}. Due to the optical mode's intensive interaction with the waveguide's top and bottom boundaries, a highly polished sapphire, as evidenced by the atomic force microscopy image in Figure \ref{fig2}(b), is employed, achieving a surface roughness average (Ra) of \changemarker{0.10} nm. \changemarker{Similarly, the top surface of the deposited SiN film exhibits a roughness of 0.13 nm.} The waveguide pattern is defined by hydrogen silsesquioxane (HSQ) ebeam resist using a Raith EBPG 5200+ electron beam lithography system with a three-pass exposure process \changemarker{with offset in both x and y axis}. To counteract \changemarker{the} charge build-up from the sapphire's insulating properties during electron beam lithography, a water-soluble polymer poly(4-styrenesulfonic acid) (PSSA) \changemarker{(300 nm)} is applied, followed by \changemarker{8 nm} gold deposition. The pattern is then etched into the SiN waveguide using a fluorine-based plasma, augmented with an increased oxygen flow to remove fluoride-carbon polymer residues on the sidewalls\cite{ji2021methods}. \changemarker{The etching mask is subsequently removed using a buffered oxide etch, and} the resultant device is shown in Figure \ref{fig2}(c). \changemarker{Next, the surfaces of both the PIC and the top sapphire piece are cleaned using RCA cleaning, followed by treatment with piranha solution. The top sapphire piece, approximately 5 mm by 5 mm in size, and the PIC surfaces are activated via oxygen plasma before being directly bonded.} Silicon oxynitride, deposited via \changemarker{plasma-enhanced chemical vapor deposition} around the non-bonded areas, serves as the top cladding with a refractive index matched to the sapphire at a visible wavelength of 780 nm. The final device is cleaved to expose facets along the sapphire's m-plane, with Figure \ref{fig2}(d) showcasing the completed chip, which houses 16 ring resonators within the bonded area. 

\begin{figure*}[h!]
\centering\includegraphics[width=1\textwidth]{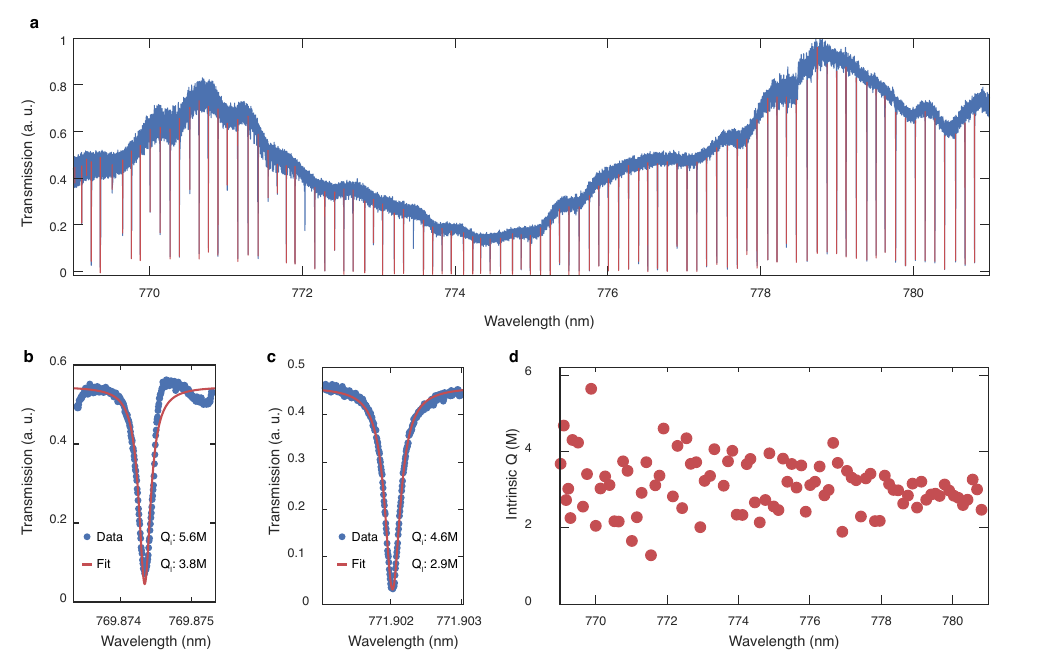}
\caption{(a) Transmission spectrum of a microring resonator, obtained through tuning a laser across the wavelength range of 769\,nm to 781\,nm. (b, c)  Zoom-in of the optical resonances at 769.874\,nm and 771.902\,nm, respectively, each measuring intrinsic quality (Q) factors of 5.6 million and 4.6 million. (d) Distribution of the intrinsic Q-factor across the near-visible spectrum.}
\label{fig3}
\end{figure*}

\section{Results}
We characterize the optical loss of \changemarker{SiN} PICs by analyzing the optical resonances at visible wavelengths, as depicted in Figure \ref{fig3}(a). A tunable laser (Newport TLB 6712-P) couples light into and out of the \changemarker{SiN} chips via lensed fibers and inverse tapers with less than 2.5 dB per waveguide facet. The transmitted light is launched into a power meter (HP81530A) and linked to a DAQ. The design of the ring resonator utilizes a point-coupling approach to mitigate loss in the coupling region. \changemarker{We observed a larger variability in resonance extinction at short wavelength, which might be attributed to coupled Fabry–Pérot cavity effects involving the bonded crystal interface and the chip facet. } 

\begin{figure*}[h!]
\centering\includegraphics[width=1\textwidth]{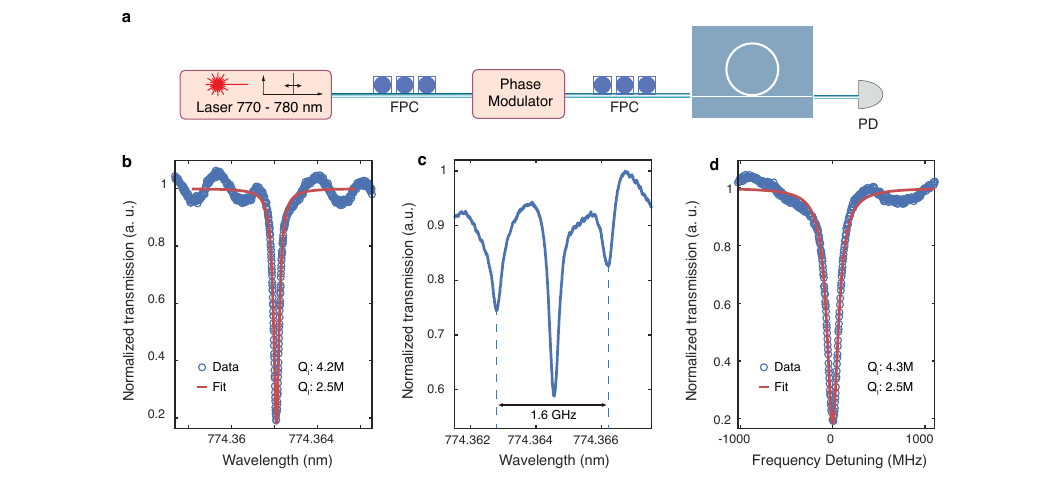}
\caption{(a) Measurement setup for evaluating resonance quality factor through optical side-band modulation. FPC: fiber polarization controller, PD: photodetector. (b) Resonance profile prior to applying the microwave drive. (c) The resonance from (b) under the influence of an 800 MHz microwave drive on the phase modulator, with a drive power of 20 dBm. (d) Calibrated resonance profile with a derived intrinsic quality factor of 4.3 million and a loaded quality factor of 2.5 million.}
\label{fig4}
\end{figure*}

\begin{table*}
    \centering
    \begin{tabular}{cccccc} 
    \toprule
         Platform&  Wavelength (nm) &  Quality factor&  Loss (dB/cm)&  Year& Ref\\ 
         \midrule 
         SiN on Sapphire&  780 &  5.6M&  0.12&  2024& This work\\ 
         SiN on Sapphire&  780 &  1.5M&  0.4&  2023& \cite{wang2023photonic}\\ 
         AlN on Sapphire&  638 &  170k&  5.3&  2018& \cite{lu2018aluminum}\\ 
         AlN on Sapphire&  369.5 &  20k&  75&  2018& \cite{lu2018aluminum}\\ 
         AlN on Sapphire&  390 &  210k&  8&  2018& \cite{liu2018ultra}\\
         AlN on Sapphire& 1550 & 2.8M& 0.13& 2019&\cite{sun2019ultrahigh}\\
         GaN/AlN on Sapphire&  1560 &  411k&  -&  2024& \cite{bhat2024gan}\\ 
         GaN on Sapphire&  1550 &  100k&  -&  2019& \cite{stassen2019high}\\ 
         GaN on Sapphire&  1550 &  2.5M&  0.17&  2022& \cite{zheng2022integrated}\\ 
         \changemarker{AlGaAs on Sapphire}&  \changemarker{1543} &  \changemarker{460k}&  \changemarker{1.2}&  \changemarker{2019}& \changemarker{\cite{zheng2019high}}\\ 
         LN on Sapphire&  1550 &  1.7M&  0.2&  2021& \cite{mishra2021mid}\\ 
         Silicon on Sapphire& 4500 & -& 4.3& 2010&\cite{baehr2010silicon}\\
          \bottomrule
    \end{tabular}
    \caption{Performance comparison of photonic waveguides on sapphire-supported platforms}
    \label{tab:comparison}
\end{table*}

The quality factor of the resonances is calculated using the equation $Q=\omega / \kappa$, where $\omega$ and $\kappa$ represent the optical frequency and the loss rate of the resonance, respectively. The mode's polarization is adjusted to TE00, which is the fundamental transverse-electric mode of the 400-$\mu$m radius microresonators. \changemarker{Figures \ref{fig3}\changemarker{(b,c)} display typical TE00 mode resonances fitted with a Lorentzian curve, with the loaded quality factors fitted to be 3.8 million and 2.9 million respectively. The extinction ratios for these resonances, reflecting the under-coupling condition ($\kappa_0 > \kappa_{\mathrm{ex}}$), are 8.95 dB and 11.6 dB, respectively. The formula used to calculate the intrinsic quality factor, $Q_{\mathrm{int}}$, from the loaded quality factor is shown below:
\[
Q_{\mathrm{int}} = \frac{2Q_{\mathrm{loaded}}}{1 \pm \sqrt{T_0}}
\]
where $T_0$ is the extinction ratio converted from decibels to a linear scale. The calculated intrinsic quality factors are 5.6 million for Figure \ref{fig3}(b) and 4.6 million for Figure \ref{fig3}(c), respectively.} Based on the Q measurements, the lowest propagation loss of the current ring resonator is derived to be 0.12 dB/cm at 769.874 nm, utilizing the formula $\alpha = 4.343\times\frac{2\pi n_g}{Q_{\mathrm{int}}\lambda}$, where $n_g=\frac{c}{2\pi R_{\mathrm{ring}}\cdot\mathrm{FSR}}$ is the group refractive index. \changemarker{The propagation losses here show a significant reduction compared to those measured in our previous publication\cite{wang2023photonic}. This improvement is attributed to two key factors: the lower surface roughness of both the sapphire substrate and the cladding, and a decrease in the waveguide thickness from 150 nm to 100 nm, which results in a weaker optical mode confinement.} Figure \ref{fig3}(d) shows a histogram of intrinsic quality factors for all fitted resonances, with a mean of 3.2 million. \changemarker{ The extinction ratios from shorter to longer wavelengths initially increase and then decrease. To ensure accuracy without overestimating the intrinsic quality factors, we converted all measurements under the assumption of under-coupling. However, it is noted that at longer wavelengths, the intrinsic $Q$ may be higher due to the over-coupling condition.}

To accurately estimate the cavity quality factor, an electro-optic modulator (EOM) is used to generate optical sidebands, serving as frequency rulers\cite{li2012sideband, zhu2024twenty}. Figure \ref{fig4}(a) shows the experimental setup, where a fiber-coupled phase modulator, driven by an RF source, generates a sinusoidally phase-modulated signal on the probe laser, producing sidebands in the frequency domain. The distance of these sidebands from the original signal offers a scale for calibrating the frequency's x-axis. Initially, a resonance was measured with phase modulation “off”, as shown in Figure \ref{fig4}(b). Subsequently, as depicted in Figure \ref{fig4}(c), two 1.6 GHz apart dips appear in the transmission spectrum, after introducing RF modulation at an 800 MHz frequency via the fiber-coupled EOM (AZ-OK5-20-PFA-PFA-UL). Utilizing these sidebands as reference points, as shown in Figure \ref{fig4}(d), the x-axis is calibrated, the linewidth ascertained, and the Q values calculated in the original resonance without the microwave drive. Through this calibrated method, we arrive at an RF-calibrated intrinsic Q value of 4.3 million, which is in close agreement with the previously established intrinsic Q value of 4.2 million.

Table \ref{tab:comparison} compares recent advancements in sapphire-based photonic platforms across various wavelengths, featuring AlN on sapphire\cite{lu2018aluminum, liu2018ultra,sun2019ultrahigh}, GaN on sapphire\cite{bhat2024gan, stassen2019high, zheng2022integrated}, LN on sapphire\cite{mishra2021mid}, \changemarker{AlGaAs on sapphire\cite{zheng2019high}},and silicon on sapphire\cite{baehr2010silicon}. As shown in Table \ref{tab:comparison}, our SiN on sapphire platform exhibits the highest intrinsic quality factor at 5.6 million and the lowest propagation loss at 0.12 dB/cm.

\section{Conclusions}
In summary, we demonstrate high-quality, visible wavelength band ring resonators featuring low-loss SiN waveguide cores, achieving an exceptional intrinsic quality factor of 5.6 million and a 35\% overlap factor with the top cladding. 
This sapphire-based approach marks a significant advance in the field of low-loss visible photonics, effectively closing a crucial gap in integration capabilities and facilitating the creation of compact, high-efficiency devices for visible photonics.

\begin{backmatter}
\bmsection{Funding}
\changemarker{Defense Advanced Research Projects Agency (HR0011-20-2-0045).} 

\bmsection{Acknowledgments}
The authors thank Michael Rooks, Yong Sun, Lauren McCabe, and Kelly Woods for their support in the cleanroom and assistance in device fabrication.

\bmsection{Disclosures}
The authors declare no conflicts of interest.

\bmsection{Data availability} \changemarker{Data underlying the results presented in this paper are not publicly available at this time but may be obtained from the authors upon reasonable request.}
\end{backmatter}

\bibliography{main_rev1}
\end{document}